\documentclass[epj]{svjour}
\usepackage{graphicx}
\usepackage{dcolumn}
\usepackage{bm}
\begin{document}
\title{Extraction of the sound velocity from rapidity spectra:\\
Evidence for QGP formation at FAIR/RHIC-BES energies}
\author{Jan Steinheimer\inst{1} \and Marcus Bleicher\inst{2,3}
}                     
%
%
\institute{ Lawrence Berkeley National Laboratory, 1 Cyclotron Road, Berkeley, CA 94720, USA \and Frankfurt Institute for Advanced Studies (FIAS), Ruth-Moufang-Str.~1, D-60438 Frankfurt am Main,
Germany \and Institut f\"ur Theoretische Physik, Goethe Universit\"at Frankfurt,\\
Max-von-Laue Str. 1, D-60438 Frankfurt Main, Germany}
\date{Received: date / Revised version: date}
%
\abstract{
We analyze longitudinal pion spectra from $\sqrt{s_{NN}}= 2$~GeV to $\sqrt
{s_{\rm NN}}=20$~GeV within Landau's hydrodynamical model and the UrQMD hybrid approach.
From the measured data on the widths of the pion rapidity spectra, 
we extract the sound velocity $c_s^2$ in the dense stage of the reactions.
It is found that the sound velocity has a local minimum (indicating a 
softest point in the equation of state, EoS) at $\sqrt{s_{NN}}= 4$-$9$~GeV, an energy range accessible at the Facility for Antiproton and Ion Research (FAIR) as well as the RHIC-Beam Energy Scan (RHIC-BES).
This softening of the EoS is compatible with the formation of a QGP at the onset of deconfinement.\\
The extracted sound velocities are then used to calculate an excitation function for the mean transverse mass of pions from the hybrid model. We find that, above $\sqrt{s_{NN}} \approx 10$ GeV, even the lowest $c_s^{2}$ gives a considerably larger $<m_T>$ of pions compared to data.
} 
\titlerunning{Extraction of the sound velocity from rapidity spectra}
\maketitle

Over the last years, a wealth of detailed data in the $E_{lab}=20A-160A$~GeV 
energy regime has become available. 
The systematic study of these data revealed surprising (non-monotonous) 
structures in various observables around $E_{lab}=30A$~GeV beam energy.
Most notable irregular structures in that energy regime include \cite{Bleicher:2011jk}, 
\begin{itemize}
\item 
the sharp maximum in the K$^+/\pi^+$ ratio \cite{Afanasiev:2002mx,Gazdzicki:2004ef,Kumar:2011us},
\item
a step in the transverse momentum excitation function (as seen through 
$\langle m_\perp\rangle -m_0$ ) \cite{Gazdzicki:2004ef,na49_blume,Kumar:2011us},
\item
an apparent change in the pion per participant ratio \cite{Gazdzicki:2004ef}
\item
a sign change in the $v_1$ slope \cite{Bedanga}
\item
an apparent maximum in the final state eccentricity from an Hanbury Brown Twiss (HBT) analysis \cite{Anson:2011ik} and
\item
a minimum in the moment products of net-proton distributions ($\sigma^2 \kappa$) \cite{Luo:2011ts}
\end{itemize}

It has been speculated, that these observation hint towards the onset of deconfinement
already at energies around $\sqrt{s_{NN}}=5$-$10$~GeV. Indeed, increased strangeness production \cite{Koch:1986ud} 
and a non-monotonous fluctuation signal have long been predicted as a sign of a mixed phase at the onset of Quark Gluon Plasma (QGP) 
formation \cite{Bleicher:2000ek,Shuryak:2000pd,Heiselberg:2000ti,Muller:2001wj,Gazdzicki:2003bb,Gorenstein:2003hk} within different 
frameworks and observables.
The suggestion of an enhanced strangeness to entropy ratio ($\sim K/\pi$) as indicator for the onset of QGP formation 
was especially advocated in \cite{SMES}. Also  the  high and approximately
constant $K^\pm$ inverse slopes of the $m_T$ spectra above $\sqrt{s_{NN}}\sim 7$~GeV - the 'step' - was also found to be consistent
with the assumption of a parton $\leftrightarrow$ hadron phase transition at low SPS 
energies \cite{Gorenstein:2003cu,Hama:2004re}.  
Surprisingly, transport simulations (supplemented by lattice QCD (lQCD) calculations) 
have also suggested that partonic degrees of freedom might already lead to
visible effects at $\sim 7$~GeV \cite{Weber98,MT-prl,Bratkovskaya:2004kv}. 
Finally, the comparison of the thermodynamic parameters $T$ and $\mu_B$
extracted from the transport models in the central overlap region
\cite{Bravina} with the experimental systematics on chemical
freeze-out configurations \cite{Braun-Munzinger:1996mq,Braun-Munzinger:1998cg,Cleymans} 
in the $T-\mu_B$ plane do also suggest that a first glimpse on a deconfined state might be possible
around  $\sqrt{s_{NN}}=5$-$10$~GeV.

\subsection{Longitudinal Dynamics}

In this paper, we explore whether similar irregularities are also present in the
excitation function of longitudinal observables, namely rapidity distributions.
Here we will employ Landau's hydrodynamical model \cite{Fermi:1950jd,Landau:gs,Belenkij:cd,Shuryak:1972zq,Carruthers:ws,Carruthers:dw,Carruthers:1981vs} and the more sophisticated UrQMD hybrid approach. 
Landau's model entered the focus again after the most remarkable observation that 
the rapidity distributions of newly produced hadrons at all investigated energies can be well 
described by a single Gaussian at each energy. The energy dependence of the rapidity width
can also be reasonably described by Landaus's model.
For recent applications of Landau's model to relativistic hadron-hadron and
nucleus-nucleus interactions the reader is referred to
\cite{Feinberg:1988et,Stachel:1989pa,Steinberg:2004vy,Murray:2004gh,Roland:2004} (and Refs. therein). Due to the simplicity of the Landau model we treat it as an approach to extract the qualitative features of the result. 

In several publications \cite{Netrakanti:2005iy,Satarov:2006jq} it has been shown that the results of Landau's model are not unambiguous and more sophisticated models yield quite different relations between the longitudinal expansion and the effective equation of state. Therefore, to complete the investigation we will also employ the UrQMD hybrid model which has been developed recently to combine the advantages of hadronic transport theory and ideal fluid dynamics \cite{Petersen:2008dd}. It uses initial conditions, generated with the UrQMD event generator \cite{Bass:1999tu,Dumitru:1999sf}, for a full 3+1D ideal fluid dynamical evolution, including the explicit propagation of the baryon current. After a transition back to the transport description, the freeze out of the system is treated dynamically within the UrQMD approach. The hybrid model has been successfully applied to describe particle yields and transverse dynamics from AGS to LHC energies \cite{Petersen:2008dd,Steinheimer:2007iy,Steinheimer:2009nn,Petersen:2010cw,Petersen:2011sb} and is therefore more reliable than the basic approach by Landau. One major advantage of the hybrid model is that, in contrary to Landaus's model, the transparency of the baryon currents is fully taken into account as done in the UrQMD pure transport calculations eliminating this uncertainty.  

\begin{figure}[t]
 \centering
\includegraphics[width=0.5\textwidth]{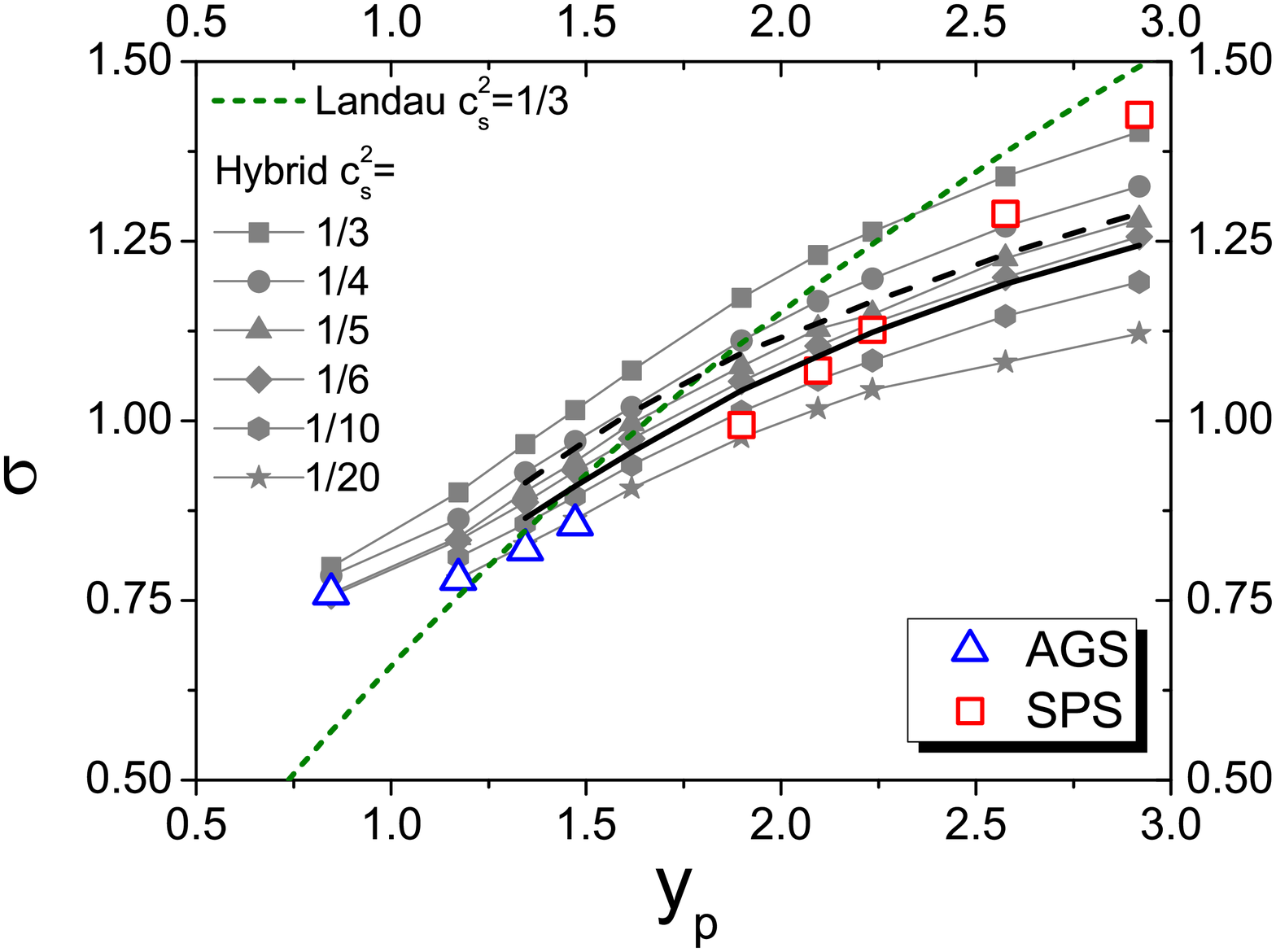} 
\caption{\label{rapwidth} The root mean square $\sigma_y$ of the rapidity
distributions of negatively charged pions in central Pb+Pb (Au+Au) reactions 
as a function of the beam rapidity $y_p$.
The green short-dashed line indicates the Landau model prediction with  $c_s^2=1/3$, while the grey lines with symbols indicate 
the UrQMD hybrid model results with different equations of state.
Data (open symbols) are taken from \cite{na49_blume,klay}.
The statistical errors given by the experiments are smaller than the symbol sizes. 
Systematic errors are not available.}
\end{figure}
The main physics assumptions of Landau's
picture are as follows: The collision of two Lorentz-contracted nuclei 
leads to full thermalization in a volume of size
$V/\gamma_{CM}$, $\gamma_{CM}$ being the Lorentz factor in the center of mass frame. This justifies the use of thermodynamics and
establishes the system size and energy dependence. For simplicity, 
chemical potentials are not taken into account in this approach.
From these assumptions follows a  universal formula for the distribution of the produced entropy, determined mainly
by the initial Lorentz contraction and a Gaussian rapidity spectrum
for newly produced particles. Further decays of resonances are not included in this approach. Under the condition that the speed of sound, $c_s$ is independent of temperature, 
the rapidity density is given by \cite{Shuryak:1972zq,Carruthers:dw}:
\begin{equation}
\frac{dN}{dy}=\frac{Ks_{\rm NN}^{1/4}}{\sqrt{2\pi \sigma_y^2}}\,\exp\left(-\frac{y^2}{2\sigma_y^2}\right)
\label{eq1}
\end{equation}
with
\begin{equation}
\sigma_y^2=\frac{8}{3}\frac{c_s^2}{1-c_s^4}\,{\rm ln}({\sqrt {s_{\rm NN}}}/{2m_p})\quad,
\label{eq2}
\end{equation}
where $K$ is a normalization factor, $m_p$ is the proton mass and $\sqrt{s_{NN}}$ the center of mass beam energy.
Under these assumptions, Landaus's model allows to relate the observed particle
multiplicity and distribution in a simple and direct analytic way to the  parameters of 
the QCD matter, namely the equation of state, under consideration.\\
For the hybrid model such a connection cannot easily be made and the width of the rapidity distributions may have non-trivial origins. In particular the initial state  plays a role, as one assumes an early system which may not be in equilibrium. Also the length and therefore importance of the fluid phase becomes larger with increasing energy. Consequently we will perform several calculations with the hybrid model, fixing the EoS to different values of the sound velocity, in the fluid dynamical phase. Then we extract the mean root square width numerically from the final state particle properties and extract the needed sound velocity from a comparison to experimental data.

\begin{figure}[t]
 \centering
\includegraphics[width=0.5\textwidth]{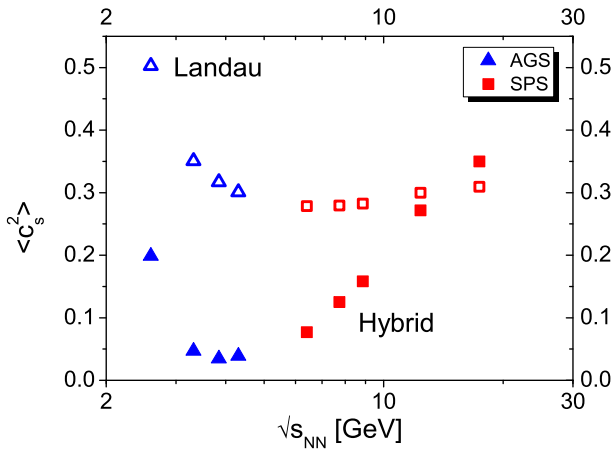} 
\caption{\label{cs2} Speed of sound as a function of $\sqrt{s_{NN}}$ for central 
Pb+Pb (Au+Au) reactions as extracted from the data using Eq.\ (\ref{eq3}) (open symbols) and from comparing to hybrid model results (full symbols).
The statistical errors (not shown) are smaller than 3\%.}
\end{figure}

As the hybrid model is a rather complex approach, it contains several parameters which can have an influence on the resulting values of the rapidity widths, e.g. the hydro-start time or the transition criterion back to the hadronic cascade. Studies on the parameters have been performed in \cite{Petersen:2008dd} and have have shown a rather small influence on results. However, further and more detailed studies, applying statistical methods which exceed the scope of this paper, are in order and are currently under way \cite{Petersen:2010zt}. One effect which might have a considerable impact on our results, the contribution of resonance decays on the pion spectra, will be discussed in the work. We will estimate the resonance contribution in the hybrid model and compare with the effect of the equation of state.  

Let us now analyze the available experimental data on rapidity distributions of negatively 
charged pions in terms of the hydrodynamical models.
Fig. \ref{rapwidth} shows the measured root mean square $\sigma_y$ of the rapidity
distribution of negatively charged pions in central Pb+Pb (Au+Au) reactions 
as a function of the beam rapidity in the center of mass frame. The green short-dashed line indicates 
the Landau model predictions with the commonly used constant sound velocity $c_s^2=1/3$. 
The grey lines with symbols depict the UrQMD hybrid model results with different equations of state (different but constant values of $c_{s}^{2}$), while the
data points \cite{na49_blume,klay} are depicted by open symbols.

Note that the black solid line in figure \ref{rapwidth} depicts results where the hybrid model is applied using an equation of state of an hadron resonance gas. When we suppress the resonance contributions to the pion spectra we obtain the dashed black line. From this comparison we can conclude that resonance decays only play a minor role in the width of the rapidity spectra. Even more, the effect is only a shift to larger widths and does not change our results qualitatively. 

At a first glance the energy dependence looks structureless.
The data seem to follow a linear dependence on the beam rapidity $y_p$ without
any irregularities.
However, the general trend of the rapidity widths is also well reproduced by 
the hydrodynamic model. 
However, there seem to be systematic deviations from an energy independent speed of sound.
At low AGS energies, the experimental points are generally
under predicted by Eq.\ (\ref{eq2}), while in the SPS energy regime Landau's model overpredicts the
widths of the rapidity distributions.
Exactly these deviations do allow to 
gain information on the equation of state 
of the matter produced in the early stage of the reaction.
For Landaus's model one can analytically invert Eq.\ (\ref{eq2}) and express the speed of sound $c_s^2$ in the medium as a function of 
the measured width of the rapidity distribution:
\begin{equation}
c_s^2=-\frac{4}{3}\frac{{\rm ln}({\sqrt {s_{\rm NN}}}/{2 m_p})}{\sigma_y^2}
+\sqrt{\left[\frac{4}{3}\frac{{\rm ln}({\sqrt {s_{\rm NN}}}/{2 m_p})}{\sigma_y^2}\right]^2+1}\quad.
\label{eq3}
\end{equation}

From the hybrid model we can extract the average speed of sound by a comparison of the experimental results with the width obtained from different equations of state. 

\begin{figure}[t]
 \centering
\includegraphics[width=0.5\textwidth]{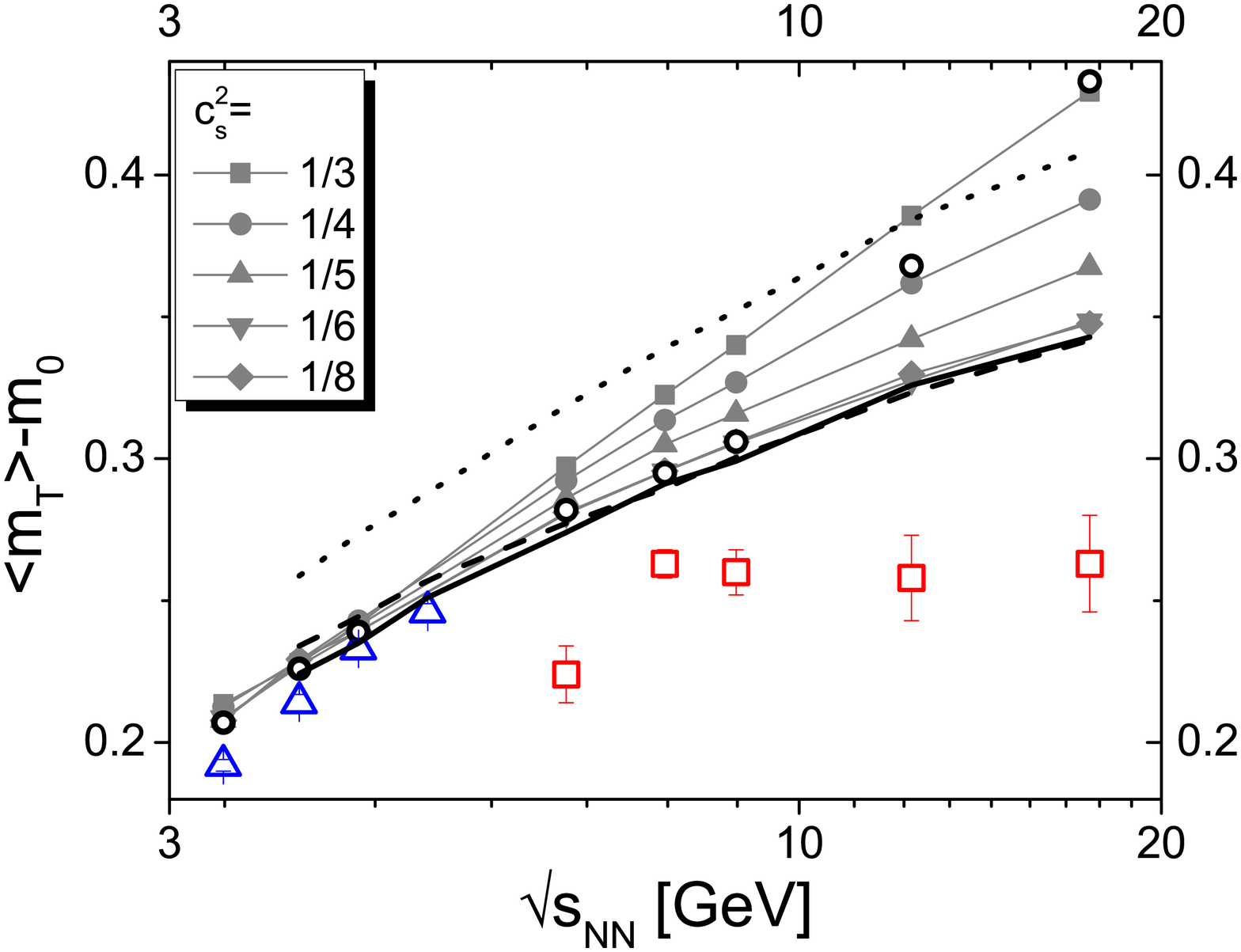} 
\caption{\label{mmt} Excitation functions of the mean transverse mass for different fixed values of the speed of sound (grey lines with symbols). The black symbols denote the mean transverse masses with speeds of sound that were extracted from the rapidity distributions in the previous section. The black lines correspond to hybrid model results with a hadron resonance gas EoS. Here we compare cases where final interactions are suppressed (dashed line) and contributions from resonance decays were omitted (short dashed line). Data are depicted as open triangle and square symbols.}
\end{figure}

Let us now investigate the energy dependence of the sound velocities extracted
from the data. Figure \ref{cs2} shows the speed of sound as a function of beam energy for central 
Pb+Pb (Au+Au) reactions as obtained from the data using Eq.\ (\ref{eq3}) and Landau's model (open symbols) as well the speed of sound
which was extracted from the full 3+1 dimensional hybrid model simulations (full symbols).
One can clearly see that the absolute values of $c_s^2$ depend considerably on the framework that is used. In general the speed of sound extracted from the 
full 3+1D hydrodynamics is smaller than from Landau's model.
On the other hand, the sound velocities for both approaches exhibit a minimum (usually called the softest point). For the Landau case it is not very pronounced and around a beam energy of $\sqrt{s_{NN}}=6$-$9$~GeV while for the hybrid model it is more pronounced and as at $\sqrt{s_{NN}}=4$~GeV.
The observed shift in the softest point can be explained by the following considerations. In Landau's model the full evolution is governed by the hydrodynamic flow, which starts to develop once the two colliding nuclei have overlapped. In particular there is no pre-equilibrium dynamics.
In the hybrid model we observe a dynamical equilibration process and strong pre-equilibrium dynamics. This is particularly important at low energies where the length 
and relevance of the fluid dynamical phase is the smallest, when compared to the initial and final decoupling phase. The extracted value of $c_{s}^2$, of the hybrid model, is therefore only sensitive on the most dense part of the collision. Therefore one should regard the two results as bounds on the energy range in which the softening of the EoS occurs.\\
\mdseries
A localized softening of the equation of state is a long predicted signal for the formation of a Quark Gluon plasma. Possible scenarios for for the observed softening include the presence of a mixed phase or a supercooled QGP in the specified energy range \cite{Hung:1994eq,Rischke:1995pe,Brachmann:1999mp,Csernai:1995zn,Csorgo:1994dd,Letessier:1993dj}. 
Therefore, we conclude that the measured data on the rapidity widths of negatively charged pions
are indeed compatible with the assumption of the onset of deconfinement in the FAIR energy range.
However, presently we can not rule out that also an increased resonance contribution may
be the cause of the softening \cite{soft}.

\subsection{Transverse Dynamics}
For a meaningful interpretation of the extracted speed of sound we relate the results found in the previous section to bulk transverse observables, in particular the mean transverse mass of pions. Fig \ref{mmt} displays the excitation functions of the mean transverse mass of charged pions from the hybrid model with different fixed values of the speed of sound (grey lines with symbols), compared to data (open symbols). The black open circles depict mean transverse masses one would obtain with speed of sound extracted in the previous section. 

From this figure it is clear that in the AGS energy range the mean transverse mass is almost independent of the applied speed of sound, while at higher energies even a very soft equations of state does not allow to obtain a sufficiently small $<m_T>$ as has been measured. To understand this behavior, figure \ref{mmt} also shows hybrid model results where we applied an EoS that resembles a hadron resonance gas (black lines). Here we compare results of the standard implementation of the hybrid model (black solid line) with results where the final cascade stage is omitted (black dashed line) and the feed down contribution of resonances to the pion mean transverse mass is omitted (black dotted line). One observes that the final cascade stage does not have any considerable influence on the mean $<m_T>$, while the resonance feed down contribution is of the same order than the dependence on the equation of state. In this context the mean transverse mass of pions seems not a well suited observable to pin down the equation of state, as one does not have full certainty on resonance contributions to the results. Therefore we advocate that the rapidity spectra might allow for a clearer investigation of the EoS due to the smaller effect of the resonance decays.

\subsection{Summary}

In conclusion, we have explored the excitation functions of the rapidity widths of negatively charged
pions in Pb+Pb (Au+Au) collisions.
\begin{itemize}
\item
The rapidity spectra of pions produced in  central nucleus-nucleus reactions at all investigated energies can be 
well described by single Gaussians.
\item
The energy dependence of the width of the pion rapidity distribution follows the
prediction of hydrodynamical models if a variation of the sound
velocity with the energy is taken into account.
\item
The speed of sound excitation function extracted from the data has a pronounced 
minimum (softest point). Although the energy at which we extracted the softest point depends on the hydrodynamical framework,
its existence turns out to be rather robust and it is located in the energy range of $E_{\rm lab}= 5A$-$30A$ GeV.\\
\item
This softest point is compatible with the formation of a Quark Gluon Plasma indicating the onset of deconfinement at this energy.
\item
Transverse observables like the mean transverse mass of pions are shown to strongly depend on the contributions of resonance decays and are therefore a more ambiguous tool to discriminate different equations of state.
\end{itemize}
Further explorations of this energy domain is needed and can be done at the FAIR and 
by CERN-SPS and BNL-RHIC experiments.

\end{document}